\documentclass[fleqn,12pt]{wlscirep}
\usepackage[utf8]{inputenc}
\usepackage[T1]{fontenc}
\usepackage[export]{adjustbox}
\usepackage[justification=justified]{caption}
\usepackage{doi}
\usepackage{caption}
\usepackage{subcaption}
\usepackage{comment}

\title{Competitive algorithms for calculating the ground state properties of Bose-Fermi mixtures}

\author[1,*]{Tomasz \'Swis\l{}ocki}
\author[2]{Krzysztof Gawryluk}
\author[2]{Miros\l{}aw Brewczyk}
\author[2,\#]{Tomasz Karpiuk}

\affil[1]{Institute of Information Technology, Warsaw University of Life Sciences -- SGGW, ul. Nowoursynowska 159, PL-02786 Warsaw, Poland}
\affil[2]{Wydzia\l{} Fizyki, Uniwersytet w Bia\l{}ymstoku, ul. K. Cio\l{}kowskiego 1L, PL-15245 Bia\l{}ystok, Poland}
\affil[*]{tomasz\_swislocki@sggw.edu.pl}
\affil[ \#]{t.karpiuk@uwb.edu.pl}

\begin{abstract}

In this work we define, analyze, and compare different numerical schemes that can be used to study the ground state properties of Bose-Fermi systems, such as mixtures of different atomic species under external forces or self-bound quantum droplets. The bosonic atoms are assumed to be condensed and are described by the generalized Gross-Pitaevskii equation. The fermionic atoms, on the other hand, are treated individually, and each atom is associated with a wave function whose evolution follows the Hartree-Fock equation. We solve such a formulated set of equations using a variety of methods, including those based on adiabatic switching of interactions and the imaginary time propagation technique combined with the Gram-Schmidt orthonormalization or the diagonalization of the Hamiltonian matrix. We show how different algorithms compete at the numerical level by studying the mixture in the range of parameters covering the formation of self-bound quantum Bose-Fermi droplets. 
\end{abstract}
\begin{document}

\flushbottom
\maketitle
%
%


\section{Introduction}

Since the first realization of Bose-Einstein condensation in vapours of ultracold atomic gases \cite{Anderson95,Davis95,Bradley95,Bradley97} and the cooling down to the degeneracy of fermionic atoms \cite{DeMarco99,Truscott01,Schreck01,Granade02,Hadzibabic03}, there has been a surge of interest in studying the quantum properties of mixtures of different atomic species. Fermionic mixtures at unitarity, mixtures in optical lattices, low-dimensional systems, or dipolar gases have attracted particular attention. A wide variety of phenomena have been observed. To mention just a few, collective oscillations \cite{Kinast04,Bartenstein04,Altmeyer07}, sound \cite{Joseph07}, and rotational modes \cite{Clancy07} have been excited in the unitary Fermi gas (and used to reveal the underlying transport parameters); strongly correlated novel magnetic phases have been induced in two-component mixtures of fermionic atoms loaded into an optical lattice \cite{Greif13,Hart15,Salomon19}; the attractive boson-boson interactions mediated by fermions have been demonstrated \cite{DeSalvo19}.

Recently, quantum droplets, self-bound systems of ultracold dilute atomic clouds, were observed in samples with large magnetic moments, such as dysprosium \cite{Schmitt16} and erbium \cite{Chomaz16} atoms. Soon after, the self-bound quantum droplets were found in a two-component mixture of bosonic potassium \cite{Cabrera18,Semeghini18} and in a heteronuclear bosonic mixture of potassium and rubidium \cite{Errico19}. In all these quantum systems, quantum corrections are essential to support self-binding. Another self-bound system, the Bose-Fermi droplets, have recently been proposed theoretically \cite{Rakshit18,Rakshit18a,Karpiuk20}. A Bose-Fermi droplet consists of a superfluid bosonic component and spin-polarized fermions. It is assumed that the system's parameters are tuned away from the $p$-wave pairing condition between spin-polarized fermions \cite{Bruun18}. Therefore, the fermions are considered to be in a normal phase. However, the properties of fully superfluid Bose-Fermi mixtures, including the formation of quantum droplets in such systems, have also been discussed \cite{Salasnich08,Adhikari18,Wang20}. 
Superconducting paired electrons (the bosonic component) embedded in unpaired electrons (the normal fermionic component) effectively form a Bose-Fermi mixture and are also an active area of research in condensed matter physics (see, for example, Ref. \cite{Kagan24}).

There are a number of algorithms for solving the dynamics of a one-component system such as a Bose-Einstein condensate (BEC), including both the short-range van der Waals and long-range dipolar interactions between atoms, in one, two, and three spatial dimensions \cite{Bao03,Muruganandam09,Bao10,Vudragovic12,Bao16,Loncar16,Kumar19,Young23,Young24}. From a numerical point of view, to find the evolution of such a system one has to solve the appropriate generalized Gross-Pitaevskii (GP) equation, and various versions of the split-step methods are used for this. Some of these algorithms can be extended to the case where BEC must actually be described by a set of coupled GP equations. This is the case, for example, when studying the spin-orbit coupled condensate or the spinor BEC in spin-1 or spin-2 hyperfine states \cite{Ravisankar21,Smith22,Banger22}. Similar techniques can be used to resolve the dynamics of a mixture of bosons and degenerate fermionic atoms. When the Fermi gas is treated as a hydrodynamic fluid, the dynamics of the Bose-Fermi mixture can be calculated from the set of two coupled nonlinear Schr\"odinger-like equations \cite{Gawryluk18}. When the fermionic gas is described in terms of the Hartree-Fock orbitals, the evolution of the Bose-Fermi mixture can still be calculated using the methods developed in Ref. \cite{Gawryluk18}. However, to find the ground state of the system, which is then manipulated and treated as the initial state for further evolution, different numerical techniques have to be employed. In this article such methods are discussed in detail.

The paper is organized as follows. We start with the description of the system in Sec.~\ref{System}, where we introduce the generalized Gross-Pitaevskii equation for bosons and a set of Hartree-Fock equations for fermions. Together they form the basic set of equations for the Bose-Fermi mixture we deal with. In Sec.~\ref{Numericalsection} we present all the numerical algorithms. 
Then, in Sec.~\ref{ResultsSection}, we compare the performance of different methods in terms of wall clock times and memory requirements. General remarks and a summary are given in Sec.~\ref{Conclusions}.

\section{Equations of motion for Bose-Fermi mixtures}\label{System}

We consider a three-dimensional Bose-Fermi mixture of alkali atoms at zero temperature. We have $N_B$ and $N_F$ bosons and fermions, respectively, with masses $m_B$ and $m_F$. We assume that the bosonic atoms interact with each other and that the interaction is repulsive. The fermionic component is spin polarized and at very low energies, which is the case at zero temperature, the interaction between fermions is forbidden. Thus, the fermions do not interact with each other. However, mutual interaction between the two components is allowed and is crucial as it couples both components. 

In principle, such a system is described by the many-body Schr\"{o}dinger equation. However, by approximating the many-body wave function by a product of the Hartree ansatz for bosons and the Slater determinant for fermions, we can simplify our modelling. Instead, we obtain a set of coupled equations, one of which is the modified Gross-Pitaevskii equation for bosons, and the other the Hartree-Fock equations for fermions. The complete set of equations is given by \cite{Rakshit18}
\begin{eqnarray}
i\hbar\frac{\partial \psi_B}{\partial t} &=& \left(-\frac{\hbar^2}{2 m_B}\nabla^2 
+ V^{B}_{ext} + \frac{\delta E_{B}}{\delta n_B} 
+ \frac{\delta E_{BF}}{\delta n_B} \right) \psi_B  \,, \label{FullSetB} \\
i\hbar\frac{\partial \psi^F_j}{\partial t} &=& \left(-\frac{\hbar^2}{2 m_F}\nabla^2
+ V^{F}_{ext}
+ \frac{\delta E_{BF}}{\delta n_F}   \right) \psi^F_j \,, 
\label{FullSet}
\end{eqnarray}
where $j = 1, \dots , N_F$. The first terms in the brackets are related to the kinetic energy. Next we have the trapping potentials for bosons and fermions. In principle, the trapping potential can have any form, but in the context of cold quantum gases it is usually the harmonic potential given below
\begin{eqnarray}
 V_{ext}^{B} &=& \frac{1}{2}m_B \left[ (\omega_x^B)^2 x^2 + (\omega_y^B)^2 y^2 + (\omega_z^B)^2 z^2 \right] \,, \nonumber \\
 V_{ext}^{F} &=& \frac{1}{2}m_F \left[ (\omega_x^F)^2 x^2 + (\omega_y^F)^2 y^2 + (\omega_z^F)^2 z^2 \right] \,,
\label{HarmonicTraps}
\end{eqnarray}
where $\omega_{\xi}^B$ and $\omega_{\xi}^F$ ($\xi=x,y,z$) are bosonic and fermionic trap frequencies, respectively. We work with spherically symmetric trap, so $\omega_{\xi}^B = \omega_B$ and $\omega_{\xi}^F = \omega_F$. Our trapping potentials can be written in the following form
\begin{eqnarray}
 V_{ext}^{B} &=& \frac{1}{2}m_B \omega_B^2 r^2 \,, \nonumber \\
 V_{ext}^{F} &=& \frac{1}{2}m_F \omega_F^2 r^2 \,.
\label{HarmonicTrapsRadial}
\end{eqnarray}
The third term in the bosonic Eq. (\ref{FullSetB}) describes the Bose-Bose interactions. The bosonic interaction energy $E_{B}$ has the following form
\begin{equation}
\label{IntEneBos} 
E_B = \int d\mathbf{r} \left( g_B\, n_B^2 /2 + C_{LHY}\, n_B^{5/2} \right) , 
\end{equation}
where $g_B = 4\pi \hbar^2 a_{B}/m_B$ is the bosonic coupling constant, $a_B$ is the bosonic scattering length, $n_B$ is the bosonic density, and $C_{LHY}=64/(15\sqrt{\pi})\,g_B\, a_B^{3/2}$. The first term in Eq. (\ref{IntEneBos}) is the usual mean-field interaction term, while the second term is the Lee-Huang-Yang quantum correction.
The fourth term in the bosonic Eq. (\ref{FullSetB}) and the third in the fermionic Eq. (\ref{FullSet}) describe the interaction between the bosonic and fermionic components, and $E_{BF}$ is given by
\begin{equation}
\label{IntEneBosFer} 
E_{BF} = \int d\mathbf{r} \left[ g_{BF}\, n_B n_F +
C_{BF}\, n_B n_F^{4/3} A(w,\alpha) \right] ,
\end{equation}
where $g_{BF} = 2\pi \hbar^2 a_{BF}/\mu$ is the mutual coupling constant, $a_{BF}$ is the scattering length corresponding to the boson-fermion interactions, $\mu=m_B\, m_F/(m_B+m_F)$ is the reduced mass, $n_F$ is the fermionic density, $w=m_B/m_F$ and $\alpha=16\pi\, n_B a_B^3 / (6\pi^2\, n_F a_B^3)^{2/3}$ are the dimensionless parameters, and $C_{BF}=(6 \pi^2)^{2/3} \hbar^2 a_{BF}^2 / 2 m_F$.
The function $A(w,\alpha)$ is given in an integral form \cite{Giorgini02}
\begin{eqnarray}
A(w,\alpha) = \frac{2(1+w)}{3w}\left(\frac{6}{\pi}\right)^{2/3}\int^{\infty}_0 {\rm d}k \int^{+1}_{-1}{\rm d}{\Omega}
 \left[ 1 -\frac{3k^2(1+w)}{\sqrt{k^2+\alpha}}
\int^{1}_0{\rm d}q q^2 \frac{1-\Theta(1-\sqrt{q^2+k^2+2kq\Omega})}{\sqrt{k^2+\alpha}+wk+2qw\Omega}  \right],
\label{AFunction}
\end{eqnarray}
where $\Theta(x)$ is the step theta function. After the functional derivatives are calculated, one gets the full form of the time-dependent equations
\begin{eqnarray}
i\hbar\frac{\partial \psi_B}{\partial t} &=& \left[-\frac{\hbar^2}{2 m_B}\nabla^2 
+ V^{B}_{ext} + g_B\, n_B
+ \frac{5}{2} C_{LHY}\, n_B^{3/2} + g_{BF}\, n_F
+  C_{BF}\, n_F^{4/3} A(\alpha) +C_{BF}\,n_B n_F^{4/3}\, \frac{\partial A}{\partial \alpha} \frac{\partial \alpha}{\partial n_B}   \right] \psi_B  \,,  
\nonumber  \\  \label{FullSetWithTermsB} \\
i\hbar\frac{\partial \psi^F_j}{\partial t} &=& \left[-\frac{\hbar^2}{2 m_F}\nabla^2
+ V^{F}_{ext}
+ g_{BF}\, n_B + \frac{4}{3} C_{BF}\, n_B\, n_F^{1/3} A(\alpha)
+ C_{BF}\,n_B\, n_F^{4/3}\, \frac{\partial A}{\partial \alpha} \frac{\partial \alpha}{\partial n_F}   \right] \psi^F_j .
\label{FullSetWithTerms}
\end{eqnarray}
The time-independent version of the set of fermionic equations is as follows
\begin{eqnarray}
\left[-\frac{\hbar^2}{2 m_F}\nabla^2
+ V^{F}_{ext}
+ g_{BF}\, n_B + \frac{4}{3} C_{BF}\, n_B\, n_F^{1/3} A(\alpha)
+ C_{BF}\,n_B\, n_F^{4/3}\, \frac{\partial A}{\partial \alpha} \frac{\partial \alpha}{\partial n_F}   \right] \psi^F_j
&=& E_j^F\psi^F_j .
\label{FullSetNoTime}
\end{eqnarray}
To find the ground state of the Bose-Fermi mixture, we use time-dependent equations for the bosons and both time-dependent and time-independent versions of the equations for the fermions. Note that quantum corrections, which appear in the expressions for the interaction energies (\ref{IntEneBos}) and (\ref{IntEneBosFer}), are essential to obtain solutions in the form of self-bound quantum Bose-Fermi droplets.

\section{Description of algorithms}
\label{Numericalsection}


The main algorithms used to find the ground state of the Bose-Fermi mixture are those based on: the adiabatic change of the mutual boson-fermion interactions during the real time propagation (A-RTP), the imaginary time propagation (ITP), and the solution of the eigenvalue problem for fermions. These algorithms are described in the following subsections. The subsection \ref{MethodsBF} discusses four methods for obtaining the ground state properties of the Bose-Fermi mixture, which are combinations of the algorithms just mentioned and will be examined in detail  in Sec.~\ref{ResultsSection}.



\subsection{Adiabatic change of mutual interactions in real time}

Here we assume that we know the ground state wave functions for the system without mutual boson-fermion interactions. This means that $g_{BF}=0$ and all terms coupling the bosonic and fermionic equations vanish. For the bosons we are still left with the interactions between bosonic atoms. The ground state for this component is easily found by the imaginary time propagation which is described in the next subsection. For fermions this means that they do not interact at all and in the harmonic trap there is a well known analytic solution
\begin{equation}
\psi_{n_x,n_y,n_z}(x,y,z) = \varphi_{n_x}(x)\varphi_{n_y}(y)\varphi_{n_z}(z) \, ,
\label{HOfunctions}
\end{equation}
where
\begin{equation}
\varphi_{n_\xi}(\xi) = \frac{1}{\sqrt{2^{n_\xi} n_{\xi}!}} 
\left( \frac{m\omega_\xi}{\pi \hbar} \right)^{1/4}
e^{-\frac{m\omega_{\xi}\xi^2}{2\hbar}}
H_{n_{\xi}}\left(\sqrt{\frac{m\omega_\xi}{\hbar}\xi}\right);
\quad\quad \xi = x,y,z \,;
\quad\quad n_{\xi} = 1, 2, 3, \dots \,.
\end{equation}
The functions $H_{n_\xi}$ are Hermite polynomials. The corresponding energy levels are
\begin{equation}
E_{n_x,n_y,n_z} = \hbar \omega_x \left(n_x + \frac{1}{2}\right) +
\hbar \omega_y \left(n_y + \frac{1}{2}\right) +
\hbar \omega_z \left(n_z + \frac{1}{2}\right).
\end{equation}
Using these solutions as the initial solution, we slowly evolve the system in time, changing the mutual coupling constant $g_{BF}$ from $0$ for $t=0$ to the desired final value $g_{BF}^{(f)}$ for $t=t_f$. We use the sine square ramp to change the coupling constant
\begin{equation}
g_{BF}(t) = g_{BF}^{(f)}\,\, \sin^2\left(\frac{\pi}{2}\, \frac{t}{t_f}\right)).
\label{Ramp}
\end{equation}
For times greater than $t_f$ the mutual coupling constant remains at the final value $g_{BF}^{(f)}$. To make the ramp adiabatic, the final time $t_f$ must be sufficiently large. The details of how the final time was chosen are given in Sec.~\ref{ResultsSection}. Some modifications to this method are still possible. For example, applying the Pitaevskii damping procedure can further smooth out the persisting tiny oscillations in the fermionic and bosonic densities (see Ref. \cite{Choi98} for details).

In real time propagation, we solve the set of coupled equations of the following type
\begin{equation}
 i \hbar \frac{\partial \psi_j}{\partial t} = \hat{H}_j \psi_j \,,
 \quad\quad j=1,2,\dots ,N_F,N_F+1 \,,
 \label{ScalarEquation}
\end{equation}
and the last orbital is just the condensate wave function, i.e. $\psi_{N_F+1}=\psi_B$. So we have $N_F$ equations for the fermions and an additional equation for the bosons. In this form each of these equations looks like a Schr\"{o}dinger equation, but one has to remember that the effective single particle operator $\hat{H}_j$ depends on $\psi_j$. This type of equation is usually called a nonlinear Schr\"{o}dinger equation. The matrix equation as given by Eqs. (\ref{ScalarEquation}) has been treated by us in Ref.~\cite{Gawryluk18}, even in the case where the effective Hamiltonian matrix remains non-diagonal. In our case, however, the Eqs. (\ref{FullSetWithTermsB}) and (\ref{FullSetWithTerms}) lead to a diagonal effective Hamiltonian matrix and consequently the set of Eqs. (\ref{ScalarEquation}) can be solved separately.

Then, to perform the numerical integration, we split the operator $\hat{H}_j$ into two usual terms
\begin{equation}
 \hat{H}_j = \hat{T}_j + \hat{V}_j \,.
\end{equation}
The first is the kinetic energy operator $\hat{T}_j = - \frac{1}{2}\nabla^2_j$ (here the physical parameters are omitted for convenience, but remember that the bosonic and fermionic atomic masses are different) and the second $\hat{V}_j$ contains all the other terms. In our case $V_j$ is just a function so the hat symbol can be omitted. A step of the time evolution according to Eqs. (\ref{ScalarEquation}) up to the first order of accuracy in $\Delta t$ can be written as follows
\begin{equation}
 \psi_j(t+\Delta t) = \exp{\left[-\frac{i}{\hbar}(\hat{T}_j + V_j)\Delta t\right]} \psi_j(t) \,,
 \quad\quad j=1,2,\dots ,N_F,N_F+1.
\end{equation}
Using the Baker-Hausdorff theorem and keeping terms up to the first order in $\Delta t$, we can write
\begin{equation}
 \psi_j(t+\Delta t) \approx \exp{\left[-\frac{i}{\hbar}\hat{T}_j \,\Delta t\right]}
 \exp{\left[-\frac{i}{\hbar} V_j \,\Delta t\right]}
 \psi_j(t) \,,
 \quad\quad j=1,2,\dots ,N_F,N_F+1.
 \label{ScalarLieTrotter}
\end{equation}
This is well known the split operator method with the Lie-Trotter splitting \cite{Lubich1}. The second order in time formula, the so called Strang splitting \cite{Lubich1}, could also be used, but does not offer any advantage in this case (see Ref. \cite{Gawryluk18}). Then we define an auxiliary function $\phi_{j}$ in the following way
\begin{equation}
 \phi_{j}(t) = \exp{\left[-\frac{i}{\hbar} V_j \,\Delta t\right]} \psi_j(t) \,.
\end{equation}
Using this function, we rewrite the Eqs. (\ref{ScalarLieTrotter}) as follows
\begin{equation}
 \psi_j(t+\Delta t) \approx \exp{\left[-\frac{i}{\hbar}\hat{T}_j \,\Delta t\right]}
 \phi_{j}(t)\,,
 \quad\quad j=1,2,\dots ,N_F,N_F+1.
\end{equation}
In the next step we go to the k-space taking the Fourier transform
\begin{equation}
 \mathcal{F}\left[\psi_j\right] \approx \mathcal{F}\left[
  \exp{\left[\frac{i\,\Delta t}{2}\nabla^2_j \right]}
 \phi_{j}(t) \right]\,,
 \quad\quad j=1,2,\dots ,N_F,N_F+1.
\end{equation}
The left hand side can be partially calculated resulting in
\begin{equation}
 \mathcal{F}\left[\psi_j\right] \approx
  \exp{\left[-\frac{i\,\Delta t}{2}k^2_j \right]}
 \mathcal{F}\left[ \phi_{j}(t) \right]\,,
 \quad\quad j=1,2,\dots ,N_F,N_F+1.
\end{equation}
Finally taking the inverse Fourier transform we are back in the coordinate space
\begin{equation}
 \psi_j(t+\Delta t) \approx
  \mathcal{F}^{-1}\left[\exp{\left[-\frac{i\,\Delta t}{2}k^2_j \right]}
 \mathcal{F}\left[ \phi_{j}(t) \right] \right]\,,
 \quad\quad j=1,2,\dots ,N_F,N_F+1.
\end{equation}
Thus, to perform the single time step, we need to compute two Fourier transforms for each equation. Technically, we do the discrete Fourier transform using the fast Fourier transform technique with the help of a specialized library called FFTW (Fastest Fourier Transform in the West \cite{FFTW}).

\subsection{Imaginary time propagation}

The imaginary time technique (ITP) is based on the substitution of the real time by a negative imaginary time, namely $t\rightarrow -i\tau$. It is a standard technique that has its origin in the properties of the solutions of the time-dependent linear Schr\"{o}dinger equation and will be briefly described below for convenience. If the Hamiltonian is time-independent the general solution can be expressed as a sum over stationary states $\varphi_j (\mathbf{r})$,
\begin{equation}
 \Psi(\mathbf{r},t) = \sum_j a_j \varphi_j (\mathbf{r}) \exp{\left(-i\frac{E_j}{\hbar}t\right)}  \,,
\label{SEGeneral}
\end{equation}
with coefficients $a_j$ determined by the initial condition,
where we assume $E_0 < E_1 \leqslant E_2 \leqslant \dots $. After transition to imaginary time Eq. (\ref{SEGeneral}) becomes
\begin{equation}
 \Psi(\mathbf{r},\tau) = \sum_j a_j \varphi_j (\mathbf{r}) \exp{\left(-\frac{E_j}{\hbar}\tau\right)}.
\label{SEGeneralImaginary}
\end{equation}
During the pure imaginary time evolution the norm is not conserved. One have to norm the solution from time to time or after each time step. It is clear from the Eq. (\ref{SEGeneralImaginary}) that the dominant term is associated with the ground state. The higher energy states become less and less important with growing energy, during the imaginary time propagation. Thus, to obtain the ground state of the system, it is sufficient to start from any function (but not perpendicular to the ground state) and perform the imaginary time propagation together with the normalization procedure as described. The length of the evolution must be sufficiently long and the time step sufficiently small. Both of these parameters affect how accurately we reach the ground state. Note, that the GP equation is nonlinear and the above arguments are not strictly applicable. Nevertheless, the procedure along the lines just described is generally accepted in the case of the GP equation.

However, here we study the Bose-Fermi mixture, the complex system consisting of bosons and many fermionic atoms. The question is how to extend the above procedure to the case where many fermionic orbitals have to be determined simultaneously and, as an additional constraint, the fermionic wave functions have to be orthogonal. We then propose to use a similar technique to find the ground state of a system described by a set of Eqs. (\ref{ScalarEquation}). In this case, we start with an arbitrary set of orthonormal functions for the fermionic orbitals and perform the imaginary time evolution for each function. During this evolution, this set of functions will lose both the norm of each function and the orthogonality property. Therefore, to find the set of functions that determine the ground state, one needs to preserve both the norm of each function and the orthogonality property. To do this, we perform the Gram-Schmidt orthonormalization procedure from time to time or after each time step. We will refer to this procedure as imaginary time propagation followed by Gram-Schmidt orthonormalization (ITP-GS).

\subsection{Iterative solution of an eigenvalue problem}
The last method is to solve the fermionic set of time-independent Eqs. (\ref{FullSetNoTime}) and is based on solving the eigenvalue problem. One can expand the $\hat{H}_j$ operator of Eq. (\ref{ScalarEquation}) in any set of orthonormal functions and obtain a matrix representation of this operator. In the linear case, if $\hat{H}_j$ does not depend on $n_F$, a single eigenvalue solution will give us the ground state of the fermionic part of the system. In our case, however, $\hat{H}_j$ depends on $n_F$, which in turn depends on $\psi_j$ for $j=1,2,\dots,N_F$. Moreover, $\hat{H}_j$ also depends on $n_B$, but for this subsection we assume that $n_B$ is fixed. In such a case, the following procedure can be used to find the ground state:

\begin{enumerate}
 \item start with an arbitrary fermionic density and compute the initial fermionic energy (including the kinetic, trapping, and interaction with bosons energy components),
 \item solve the eigenvalue problem (Eqs. (\ref{FullSetNoTime})) and get new $\psi_j$ and $E^F_j$,
 \item calculate new fermionic density $n_F=\sum_{j=1}^{N_F}|\psi_j|^2$,
 \item compute new fermionic energy and compare with previous one, if convergence criterion is met we are done, if not repeat from point 2.
\end{enumerate}
We will refer to this procedure as the iterative eigenvalue problem (IEV).

\subsection{Methods for obtaining the ground state of the Bose-Fermi mixture}
\label{MethodsBF}

The first method uses ITP for the bosonic equation and IEV for the fermionic equations. In addition, in IEV we use the basis of harmonic oscillator wave functions (\ref{HOfunctions}) in three-dimensional form. This means that each function is stored in three-dimensional form. So we don't have to multiply one-dimensional harmonic functions for each spatial step. This approach is faster than using one-dimensional functions, but it consumes a lot of random access memory (RAM). The algorithm is as follows:

\begin{enumerate}
 \item start at any $n_B$, $n_F$ and calculate the total initial energy,
 \item for fixed $n_B$ do IEV for fermions:
 \begin{enumerate}
  \item solve the eigenvalue problem and get new $\psi_j$ and $E^F_j$,
  \item calculate the new fermionic density $n_F=\sum_{j=1}^{N_F}|\psi_j|^2$,
  \item calculate the new fermionic energy and compare it with the previous one, if the convergence criterion is met go to point 3, if not repeat from point 2(a),
 \end{enumerate}
 \item for fixed $n_F$ perform the ITP for the bosonic equation:
 \begin{enumerate}
  \item do the ITP for a fixed number of time steps
  \item compute the new bosonic energy and compare it with the previous one, if the convergence criterion is met go to point 4, if not repeat from point 3(a),
 \end{enumerate}
 \item calculate the new total energy and compare it with the energy from the previous step, if the convergence criterion is met the iteration procedure ends, if not go to step 2.
\end{enumerate}
We will use the abbreviation ITP-IEV-3D for this method.

In the second method we use one-dimensional harmonic wave functions. We are solving a three-dimensional problem. This means that every time we need the three-dimensional wave function, we multiply three one-dimensional wave functions. This time we have to store one-dimensional wave functions in memory. This way the memory footprint is much smaller. It also means that the size of the basis can be much larger. Unfortunately, we have to do more calculations. The algorithm itself is exactly the same as the first one. We call this method ITP-IEV-1D.

The third method is based purely on the adiabatic change of the mutual coupling constant during real-time propagation. 
We start from a system without mutual interactions. The internal boson interactions are already present. The initial bosonic function is obtained from the ITP for bosons. The initial fermionic orbitals are noninteracting harmonic oscillator solutions.
Both the bosonic Eq. (\ref{FullSetWithTermsB}) and the set of fermionic Eqs. (\ref{FullSetWithTerms}) are propagated in real time. The mutual interaction is ramped from $0$ to the final value according to Eq. (\ref{Ramp}). We use the name A-RTP for this method.

In the last method we use ITP for bosons and ITP-GS for fermions. The full set of Eqs. (\ref{FullSetWithTermsB},\ref{FullSetWithTerms}) is propagated in imaginary time for a certain number of time steps. Then the total energy is calculated and compared with the previous one. If the convergence criterion is met, we assume that the ground state has been found. If not, the propagation continues. We call this method ITP-ITP-GS.

\subsection{Parallelization}
\label{parallelization}

Our algorithms and the methods based on the above algorithms are parallelized for systems with shared memory. In such a system, all processors or cores have access to shared RAM. We use the Open Multi-Processing (OpenMP) standard. Typically, a shared memory system is a single workstation or one node of a cluster of nodes. Nowadays, it means that our programs can run on anywhere from one to several hundred cores in parallel.

When we perform a time step of RTP or ITP propagation of the bosonic equation, all operations performed over the three-dimensional grid are parallelized. The FFTW library also allows the parallelized version of the FFT to be used.
In the case of RTP or ITP propagation of fermionic equations, it is better to run each equation describing a fermionic orbital in parallel. Thus, for RTP, there should be no significant advantage to running the program on more cores than the number of fermionic orbitals considered. However, in the case of ITP-GS, where the GS procedure is performed after every time step or after a few time steps and the parallelization is done over a spatial grid, it should still be advantageous to use more cores than the number of fermionic orbitals.

For methods based on IEV, the size of the basis ranges from a few hundred to thousands of functions. The program spends most of its time computing the Hamiltonian and then solving the eigenvalue problem. Both steps are amenable to parallelization, and since the size of the basis is usually over hundreds of functions, parallelization should work well even on tens of cores.

\section{Numerical results}
\label{ResultsSection}

In this section we study the efficiency of four different methods, presented in Sec.~\ref{MethodsBF}, for calculating the ground state of the Bose-Fermi mixture. The system consists of $N_B$ bosons (${}^{133}$Cs atoms) and $N_F$ fermions (${}^{6}$Li atoms), which are confined in spherically symmetric traps. The trap frequencies for bosons and fermions are set to $\omega_r^{B} = 2 \pi \times 1200$Hz and $\omega_r^{F} = 2 \pi \times 4800$Hz, respectively. The attractive boson-fermion interaction parameter varies between $g_{BF} = 0$ and $g_{BF} = -5$ in units of $g_B$. We use a $128\times 128 \times 128$ spatial grid with a spatial step of $\Delta x = \Delta y = \Delta z = 3$ in units of $a_B$. The time step is equal to $\Delta t = 0.05$ (ITP-ITP-GS and A-RTP methods) and $\Delta t = 0.01$ (ITP-IEV-1D and ITP-IEV-3D methods), in units of $m_B a_B^2/\hbar$.

To compare results between different algorithms, we perform numerical simulations for the system described in Sec.~\ref{System} for different values of the mutual interaction strength $g_{BF}$, typically for $g_{BF} = -1$ and $g_{BF} = -3$. The calculations were performed on a computer with 96 GB of RAM and two Intel(R) Xeon(R) CPU E5-2650 v2 processors, each equipped with 8 cores (16 cores in total) and hyper-threading technology enabled. Compiler options are listed in the appendix \ref{compiler}.
For all methods we require an energy accuracy of the order of $10^{-7}$ (in units of $\hbar^2/(m_B a_B^2)$) when calculating the total energy. This means that the simulations are stopped when the energy difference calculated for the successive moments separated by (in our case) $2\times 10^3$ time steps is below $10^{-7}$, and also that further tuning of the method parameters (i.e. decreasing the time step) changes the energy only at the level below $10^{-7}$. Under such a condition it turns out that the ITP-ITP-GS algorithm gives the lowest value for the energy of the Bose-Fermi system when searching for the ground state (see Tab.~\ref{tab:energies}). Then the ITP-ITP-GS method will be the reference method in our case.

\begin{figure}[!thb]
\begin{center} 
  \includegraphics[width=8.0cm]{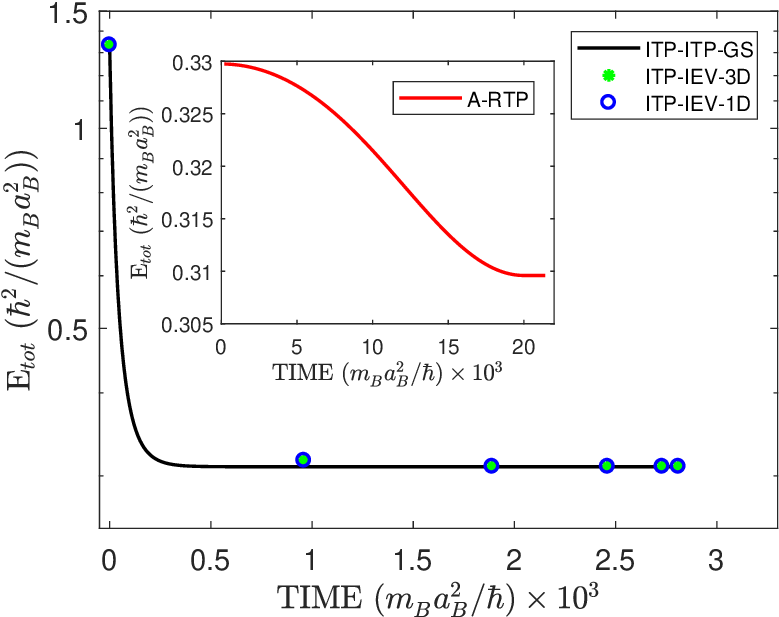}  \hspace{1cm}
  \includegraphics[width=7.9cm]{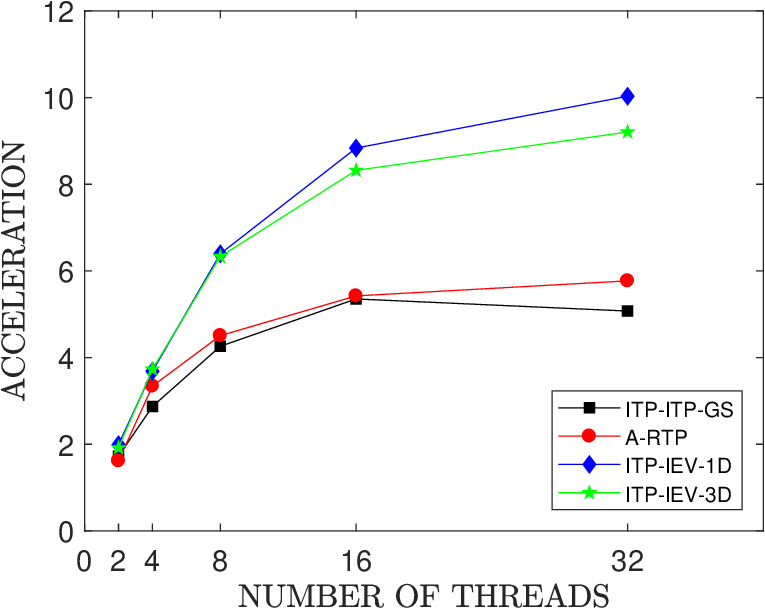}  
\end{center}  
  \caption{(Left) Total energy (in units of $\hbar^2/(m_B a_B^2)$) of a trapped Bose-Fermi mixture as a function of imaginary time (or real time for the inset), in units of $m_B a_B^2/\hbar$, for $g_{BF}=-1$ for all studied algorithms. The main frame shows the energy for the ITP-ITP-GS (black solid line), ITP-IEV-3D (green dots), and ITP-IEV-1D (blue circles) methods, while the inset depicts the energy in the case of the A-RTP algorithm. 
 (Right) Simulation acceleration (which is the inverse of the ratio of the wall clock times for the current and single-core computations)  as a function of the number of threads. The acceleration is better for the ITP-IEV-1D and ITP-IEV-3D methods.}
  \label{fig:entot_dt001_g1}
\end{figure}

We then study the properties of the system by looking at the bosonic and fermionic densities. For such high energy accuracy, the maximum change in atomic density over the entire numerical grid remains below $10^{-8}$ (in units of $1/a_B^3$) for the ITP-ITP-GS method. Obviously, results with the required accuracy are obtained with different parameters for different methods. For example, to satisfy the energy-based criterion mentioned above, the iterative and Gram-Schmidt orthonormalization methods require an order of magnitude shorter time than the algorithm that changes the boson-fermion interaction parameter in time, see Fig.~\ref{fig:entot_dt001_g1}. In addition, to keep the density changes at the same level for all methods, the time step for the ITP-IEV-1D and ITP-IEV-3D cases should be further reduced (compared to that required to satisfy the energy criterion). Nevertheless, maintaining the energy and density based criteria of simulation accuracy allows us to compare the time and space complexity of different methods.

Figure \ref{fig:entot_dt001_g1} (left frame) depicts the total energy of the Bose-Fermi mixture as a function of imaginary or real time for all considered methods. Although the rate of change may be different, the final energies, i.e. the ground state energy of the Bose-Fermi mixture, remain to be very close. The Tab.~\ref{tab:energies} helps to distinguish between the various methods. It is clear that the differences are $\lesssim 10^{-3}$ both for the system consisting of $N_B = 40$ bosons and $N_F = 4$ fermions (second and third columns) and for the system consisting of $N_B = 100$ bosons and $N_F = 10$ fermions (last two columns). Note that the final energy in the case of the ITP-IEV-3D and ITP-IEV-1D methods exhibits some differences both for weaker and stronger boson-fermion attraction and for smaller and larger systems, due to the available RAM limitation as discussed in Sec.~\ref{convergence}. Assuming that the same set of basis eigenvectors is used, both the ITP-IEV-3D and ITP-IEV-1D methods should give the same results, just by their construction. There are only a few points marked in Fig.~\ref{fig:entot_dt001_g1} (left frame) for these methods, because the total energy of the Bose-Fermi mixture is calculated here only after the bosonic part of the energy satisfies the convergence criterion (Sec.~\ref{MethodsBF}), which is numerically defined as the energy change (for moments separated by $2\times 10^3$ time steps) being below $10^{-7}$, i.e. in the same way as for the total energy of the mixture. Indeed, the ITP-ITP-GS reference method leads to the lowest value of the ground state energy of the Bose-Fermi mixture (see Tab.~\ref{tab:energies}).
 
\begin{table}[!htb]
  \begin{center}
    \caption{Ground state energies (in units of $\hbar^2/(m_B a_B^2)$) obtained by different algorithms for $g_{BF} = -1$ and $g_{BF} = -3$ according to the energy-based criterion. The system consists of $N_B = 40$ bosons and $N_F = 4$ fermions (the second and third columns) or $N_B = 100$ bosons and $N_F = 10$ fermions (the last two columns). }
    \label{tab:energies}
    \begin{tabular}{c|c|c|c|c} 
      \textbf{Algorithm} & \textbf{$E_{tot}$($g_{BF} = -1$}) & \textbf{$E_{tot}$($g_{BF} = -3$)} & \textbf{$E_{tot}$($g_{BF} = -1$}) & \textbf{$E_{tot}$($g_{BF} = -3$)} \\
      \hline
      ITP-IEV-3D & 0.3098283 & 0.1996210 & 0.9702401 & 0.5073803 \\
      ITP-IEV-1D & 0.3096138 & 0.1991338 & 0.9691752 & 0.5066022 \\
      A-RTP & 0.3095978 & 0.1991274 & 0.9689032 & 0.5062738      \\ 
      ITP-ITP-GS & 0.3095975 & 0.1991070 & 0.9689016 & 0.5061495 \\  
      \hline
      System & \multicolumn{2}{c|}{$N_B=40,\, N_F=4$} & \multicolumn{2}{c}{$N_B=100,\, N_F=10$}      
    \end{tabular}
  \end{center}
\end{table}

\begin{figure}[!htb]
\centering
\begin{subfigure}{.5\textwidth}
  \centering
  \includegraphics[width=.9\linewidth]{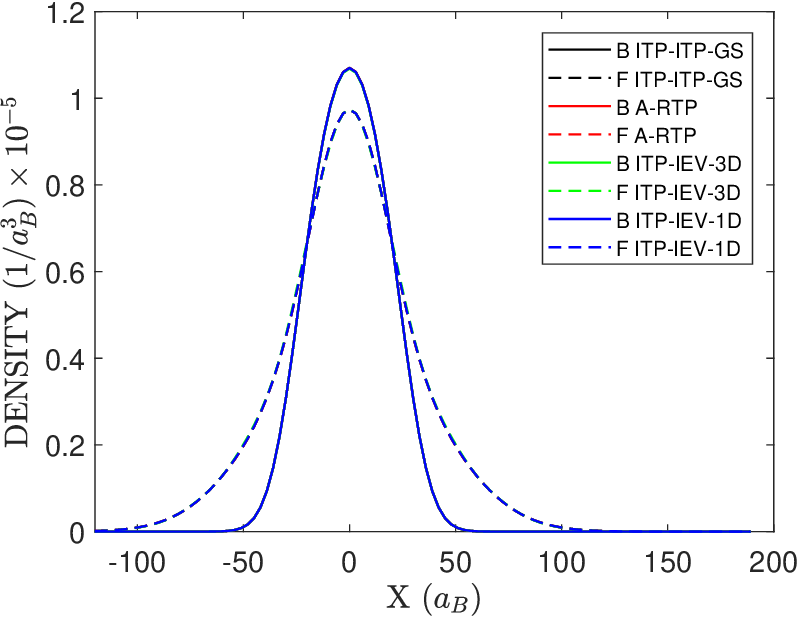}
\end{subfigure}%
\begin{subfigure}{.5\textwidth}
  \centering
  \includegraphics[width=.9\linewidth]{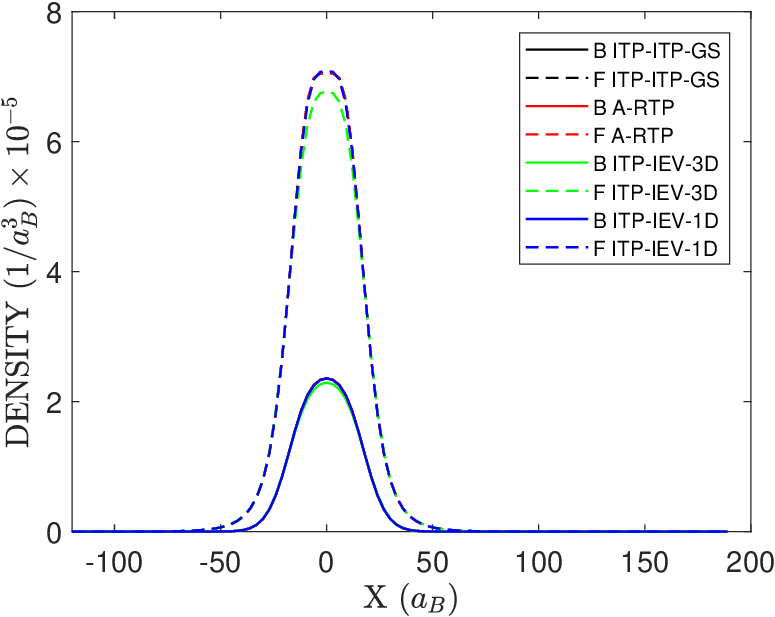}
\end{subfigure}
\caption{Radial bosonic (solid lines) and fermionic (dashed lines) densities for different methods for $g_{BF} = -1$ (left) and $g_{BF} = -3$ (right). The system consists of $N_B = 40$ bosons and $N_F = 4$ fermions and the densities are normalized to one and $N_F = 4$ for bosons and fermions, respectively. When comparing the ITP-IEV-3D method with others for $g_{BF} = -1$, no differences in the density profiles are visually visible. For higher attractive interactions, $g_{BF} = -3$, the situation changes, small differences in the density profiles are visible due to the insufficient number of eigenvectors used in the ITP-IEV-3D method.}
\label{fig:dens_g1_g3_all}
\end{figure}

\begin{figure}[!htb]
\centering
\begin{subfigure}{.5\textwidth}
  \centering
  \includegraphics[width=.9\linewidth]{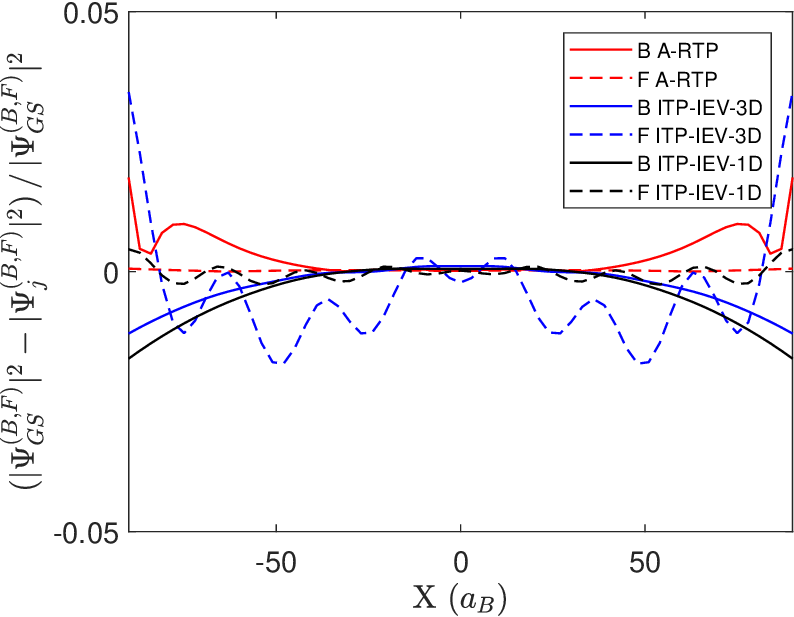}
\end{subfigure}%
\begin{subfigure}{.5\textwidth}
  \centering
  \includegraphics[width=.9\linewidth]{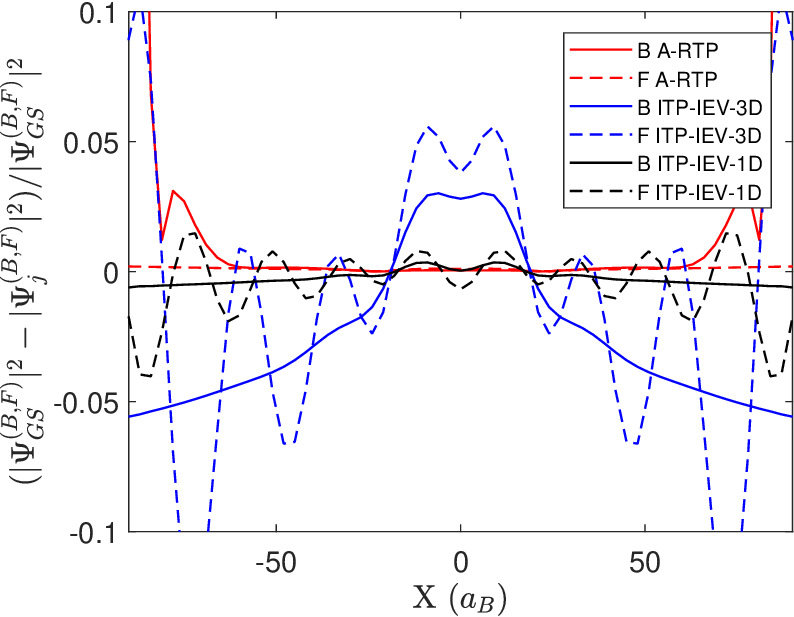}
\end{subfigure}
\caption{Relative density differences ($(|\psi_{j}^{(B,F)}|^2 - |\psi_{GS}^{(B,F)}|^2) / |\psi_{GS}^{(B,F)}|^2$) for different methods for $g_{BF} = -1$ (left) and $g_{BF} = -3$ (right). The ``j'' denotes the A-RTP, ITP-IEV-3D, and ITP-IEV-1D algorithms, while ``GS'' is the reference ITP-ITP-GS method. Solid and dashed lines correspond to fermionic and bosonic densities, respectively.}
\label{fig:dens_diff_all_g13_v2}
\end{figure}

In Fig.~\ref{fig:dens_g1_g3_all} we show the bosonic and fermionic density profiles for all considered methods for $N_B = 40$ bosons and $N_F = 4$ fermions. For small boson-fermion attraction ($g_{BF} = -1$) the differences in the density profiles are minimal. However, in the case of $g_{BF} = -3$ the peak density for the ITP-IEV-3D method differs from all the others. This is due to the fact that the accuracy of the ITP-IEV-3D method is limited by the amount of RAM available, since all the three-dimensional basis vectors used must be kept in memory at all times. For $g_{BF} = -3$ a large number of basis eigenvectors is required. Due to memory limitations, this number has to be reduced in the ITP-IEV-3D method compared to the ITP-IEV-1D method, which results in the changes of the densities (Fig. \ref{fig:dens_g1_g3_all}, right frame) and of the ground state energy (Tab. \ref{tab:energies}). The differences in the densities are more visible when we plot the relative density changes with respect to those obtained by the Gram-Schmidt orthonormalization technique (Fig. \ref{fig:dens_diff_all_g13_v2}). In Tab. \ref{tab:table3} we present the wall clock (elapsed real) times for simulations using different methods for $g_{BF} = -1$ and $g_{BF} = -3$ both for ($N_B = 40$, $N_F = 4$) and ($N_B = 100$, $N_F = 10$) systems. It is clearly visible that the ITP-ITP-GS method not only gives the lowest total energy, but also turns out to be the most efficient. Note also that the elapsed time for all the methods considered, except the reference one, is very sensitive to the strength of the Bose-Fermi interactions.

\begin{table}[!htb]
  \begin{center}
    \caption{Comparison of wall clock times (in units of hours) for simulations using different algorithms. The time given is the minimum time required to satisfy the energy-based (accuracy in total energy which is of the order of $10^{-7}$) and density-based (accuracy in densities which is of the order of $10^{-8}$) criteria. The system consists of $N_B = 40$ bosons and $N_F = 4$ fermions (second and third columns) or $N_B = 100$ bosons and $N_F = 10$ fermions (last two columns). The simulations are run on $16$ cores ($32$ threads). }
    \label{tab:table3}
    \begin{tabular}{c|c|c|c|c} 
      \textbf{Algorithm} & \textbf{Time} &  \textbf{Time} & \textbf{Time} &  \textbf{Time} \\
      & \textbf{($g_{BF} = -1$)} &  \textbf{($g_{BF} = -3$)} & \textbf{($g_{BF} = -1$)} &  \textbf{($g_{BF} = -3$)} \\
      \hline
      ITP-IEV-3D & 10.1h & 41.8h & 11.9h & 36.2h \\
      ITP-IEV-1D & 14.4h & 58.9h & 16.5h & 58.2h \\      
      A-RTP & 22.2h & 25.5h & 42.4h & 65.5h \\ 
      ITP-ITP-GS & 1.9h & 2.0h & 5.4h & 6.2h \\
      \hline
      System & \multicolumn{2}{c|}{$N_B=40,\, N_F=4$} & \multicolumn{2}{c}{$N_B=100,\, N_F=10$}   
    \end{tabular}
  \end{center}
\end{table}

In Fig. \ref{fig:entot_dt001_g1} (right frame) we show the dependence of the simulation acceleration (which is the inverse of the ratio of the wall clock times for the current and the single-core computations) on the number of threads for the methods discussed in this paper. In all cases, the simulation acceleration saturates at a certain number of cores. However, the ITP-IEV-1D and ITP-IEV-3D methods appear to be better from a parallelization point of view. This is consistent with the description given in Sec.~\ref{parallelization}. Now, some results related to the simulation times (Tab.~\ref{tab:table3}) can be better understood. The wall clock time for the A-RTP method is about $10$ times longer compared to the most efficient ITP-ITP-GS method (Tab.~\ref{tab:table3}). Since the parallelization affects both methods in the same way (see Fig.~\ref{fig:entot_dt001_g1}, right frame), the difference comes from the fact that for the A-RTP algorithm the ramp must be long enough to reach the ground state of the mixture. To do this, the A-RTP requires $10$ times more numerical steps to be performed (Fig.~\ref{fig:entot_dt001_g1}, left frame), then the elapsed time turns out to be $10$ times longer. This is especially true for the systems consisting of a smaller number of fermions, since then the Gram-Schmidt orthonormalization procedure (which itself is also parallelized) does not take much time. The wall clock times for the ITP-IEV-1D method are systematically longer than for the ITP-IEV-3D method. This is because both approaches are similarly sensitive to parallelization (Fig. \ref{fig:entot_dt001_g1}, right frame), and in the ITP-IEV-1D case an additional task must be performed to generate three-dimensional basis vectors at each time step. Then, assuming the same eigenvector basis is used, the wall clock time for the ITP-IEV-1D must be longer. Surprisingly, the wall clock time for stronger attraction, $g_{BF}=-3$, becomes shorter for larger systems. This is an artifact of the method related to the size (which is not large enough) of the basis used. We have checked that for a larger system, consisting of $N_B=200$ bosonic and $N_F=20$ fermionic atoms, the elapsed time for the ITP-IEV-3D method gets even shorter and equals $29.3$h.

\subsection{Convergence of numerical results}
\label{convergence}

\begin{figure}
\centering
\begin{subfigure}{.5\textwidth}
  \centering
  \includegraphics[width=.9\linewidth]{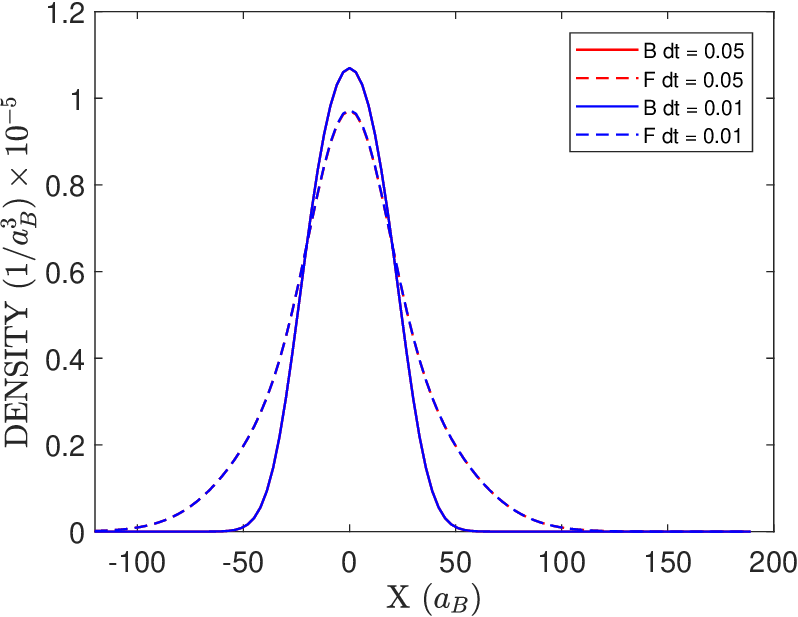}
\end{subfigure}%
\begin{subfigure}{.5\textwidth}
  \centering
  \includegraphics[width=.9\linewidth]{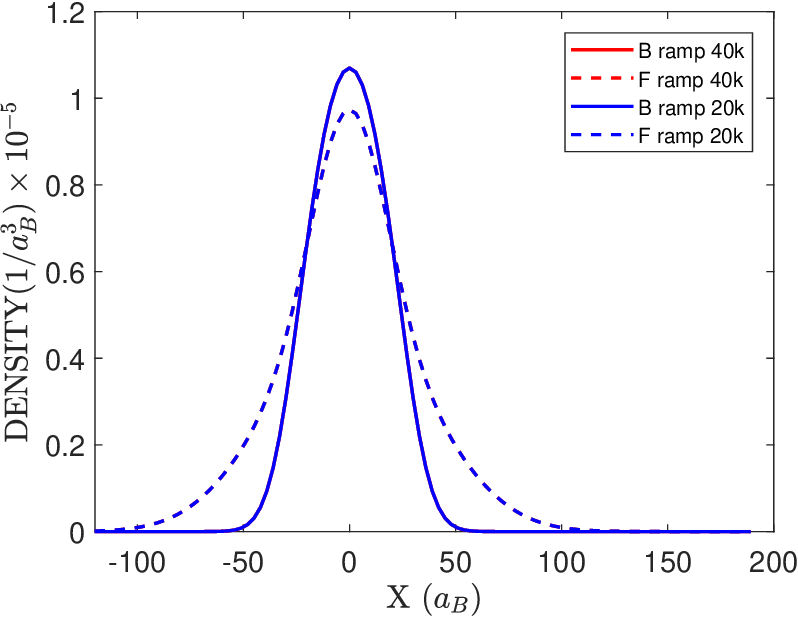}
\end{subfigure}
\caption{(Left) Densities of the system for the ITP-ITP-GS algorithm for different time steps. The solid (dashed) lines denote the fermionic (bosonic) density. Minimal differences can be seen between the red and blue lines. A time step smaller than $\Delta t = 0.01$ does not improve the density profiles. The interaction parameter is equal to $g_{BF} = -1$. (Right) Densities of the system for the A-RTP algorithm for different ramps. The time step is $\Delta t = 0.05$ and the interaction parameter is $g_{BF} = -1$. 
Red lines show $40\times 10^3$ time unit ramp while blue lines are $20\times 10^3$ time unit ramp. Lower time step or longer ramp does not improve the density profiles.}
\label{fig:DensitiesVsTimeStep}
\end{figure}

All numerical results presented above have been obtained with the best possible choice of parameter set. Now we discuss how the convergence of the considered methods depends on the numerical parameters. Different approaches are characterized by different parameter sets, which include the spatial step, the time step, the shape of the ramp and its duration (in the case of the A-RTP method), and the size of the eigenvector basis (in the case of the iterative methods). All simulations were run on the same spatial grid. The time step for each method has been chosen in such a way that the total energy of the Bose-Fermi mixture is determined at the level below $10^{-7}$ and simultaneously the change of the bosonic and fermionic densities over the entire numerical grid remains at the level below $10^{-8}$. Fig. \ref{fig:DensitiesVsTimeStep} (left panel) proves that reducing the time step from $\Delta t = 0.05$ to $\Delta t = 0.01$ does not improve the density profiles for the ITP-ITP-GS algorithm. In fact, the time step $\Delta t = 0.05$ was used to obtain all the reference (i.e., ITP-ITP-GS method-related) values reported in this paper. For the A-RTP method, both the time step and the duration of the ramp are important. Fig. \ref{fig:DensitiesVsTimeStep} (right panel) shows that the change (according to the sine-squared ramp, see Eq. \ref{Ramp}) of the interaction parameter in time $2\times 10^4\, m_B a_B^2/\hbar$ is long enough to give a reasonable approximation to the ground state bosonic and fermionic densities. Therefore, such a ramp is used throughout this paper. Note, however, that the total energy of the Bose-Fermi mixture is still slightly higher than that obtained in the ITP-ITP-GS method (see Tab.~\ref{tab:energies}), leaving room for improvement in the shape of the ramp.

\begin{figure}[!thb]
\centering
\begin{subfigure}{.5\textwidth}
  \centering
  \includegraphics[width=.9\linewidth]{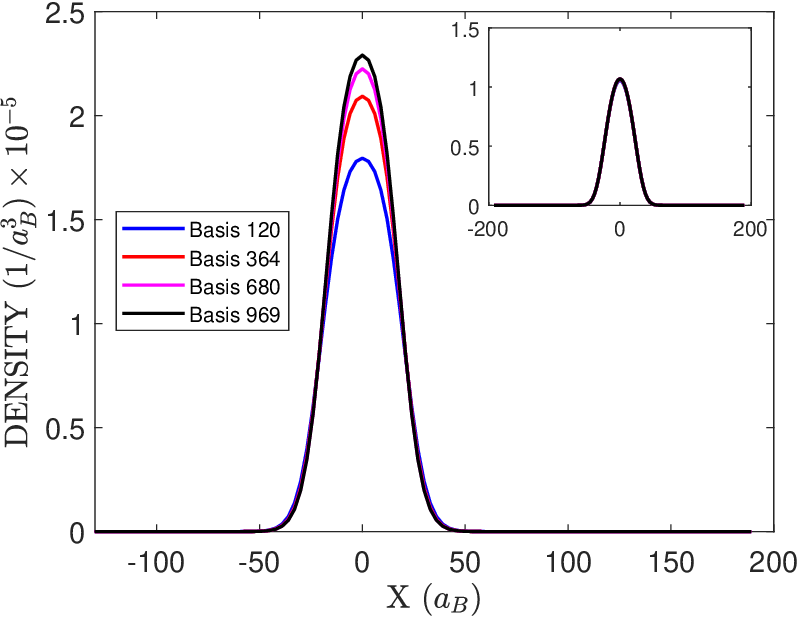}
\end{subfigure}%
\begin{subfigure}{.5\textwidth}
  \centering
  \includegraphics[width=.9\linewidth]{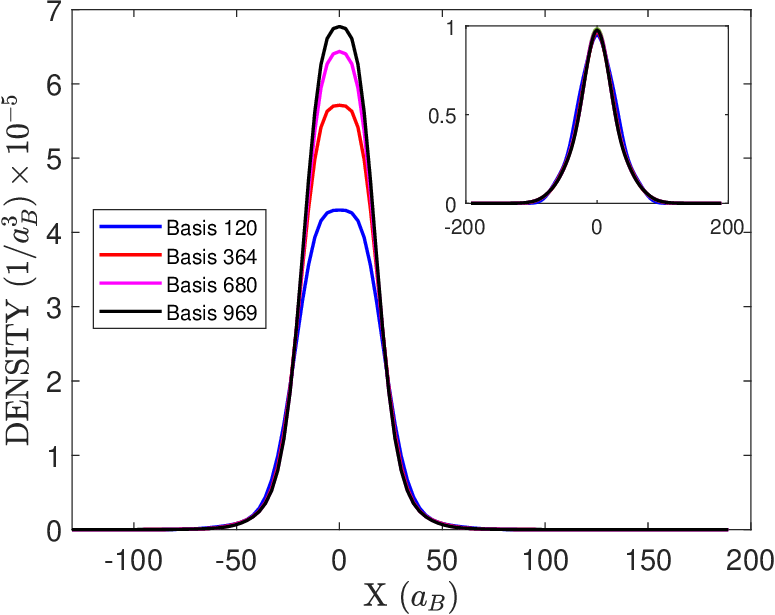}
\end{subfigure}
\caption{Results for the ITP-IEV-3D method for different numbers of basis eigenvectors for $g_{BF} = -3$ (main panels) and $g_{BF} = -1$ (insets). (Left) Densities of the bosons. (Right) Densities of the fermions. Different colors indicate different numbers of basis eigenvectors. The system consists of $N_B = 40$ bosons and $N_F = 4$ fermions.}
\label{fig:ITP-IEV-3D}
\end{figure}

\begin{table}[!htb]
  \begin{center}
    \caption{Wall clock times and RAM consumption for the ITP-IEV-3D method for $g_{BF} = -1$ and $g_{BF} = -3$ and using different eigenvector bases. The system consists of $N_B = 40$ bosons and $N_F = 4$ fermions (the third and fourth columns) or $N_B = 100$ bosons and $N_F = 10$ fermions (the last two columns). According to Fig. \ref{fig:ITP-IEV-3D}, even the largest eigenvector basis is still insufficient, since the saturation (for $g_{BF} = -3$) is continuous. }
    \label{tab:table1}
    \begin{tabular}{c|c|c|c|c|c} 
      \textbf{Number of basis} & \textbf{RAM} & \textbf{Time} & \textbf{Time} & \textbf{Time} & \textbf{Time} \\
      \textbf{eigenvectors} & & ($g_{BF} = -1$) & ($g_{BF} = -3$) & ($g_{BF} = -1$) & ($g_{BF} = -3$) \\
      \hline
      120 & 7.9\,GB  & 4.6h  & 12.2h  & 5.1h  & 8.1h   \\
      364 & 23.4\,GB & 5.4h  & 17.7h  & 5.4h  & 14.1h  \\
      680 & 43.6\,GB & 7.2h  & 23h    & 7.8h  & 24.8h  \\ 
      969 & 63.4\,GB & 10.1h & 41.1h  & 11.9h & 36.2h  \\
      \hline
      \multicolumn{2}{c|}{System} & \multicolumn{2}{c|}{$N_B=40,\, N_F=4$} & \multicolumn{2}{c}{$N_B=100,\, N_F=10$}      
    \end{tabular}
  \end{center}
\end{table}

\begin{table}[!htb]
  \begin{center}
    \caption{Ground state energies for the ITP-IEV-3D method for $g_{BF} = -1$ and $g_{BF} = -3$ and using different eigenvector bases. The system consists of $N_B = 40$ bosons and $N_F = 4$ fermions (the second and third columns) or $N_B = 100$ bosons and $N_F = 10$ fermions (the last two columns). }
    \label{tab:tableEnTotVsBase}
    \begin{tabular}{c|c|c|c|c} 
      \textbf{Number of basis} & \textbf{Total energy} & \textbf{Total energy} & \textbf{Total energy} & \textbf{Total energy} \\
      \textbf{eigenvectors} & ($g_{BF} = -1$) & ($g_{BF} = -3$) & ($g_{BF} = -1$) & ($g_{BF} = -3$) \\
      \hline
      120 & 0.3151634 & 0.2079778    &        1.0068943 & 0.5538557 \\
      364 & 0.3108000 & 0.2015291    &        0.9798275 & 0.5195266 \\
      680 & 0.3101370 & 0.2000821    &        0.9717417 & 0.5089908 \\ 
      969 & 0.3098283 & 0.1996210    &        0.9702401 & 0.5073803 \\    
      \hline
      System & \multicolumn{2}{c|}{$N_B=40,\, N_F=4$} & \multicolumn{2}{c}{$N_B=100,\, N_F=10$}   
    \end{tabular}
  \end{center}
\end{table}

In the case of the iterative methods ITP-IEV-1D and ITP-IEV-3D we need to use a large enough number of basis eigenvectors to get satisfactory results (compare Tabs. \ref{tab:tableEnTotVsBase} and \ref{tab:energies}). However, even for a system with a small number of fermions, a large number of basis eigenvectors is required, and this means a large use of RAM in the case of ITP-IEV-3D (see Tab.~\ref{tab:table1}). The maximum size of the three-dimensional eigenvector basis that can be loaded into RAM in our case is $969$. In contrast to the ITP-IEV-1D method, the ITP-IEV-3D method is limited by the RAM size of the computer. Obviously, the results are better for larger eigenvector bases, since the densities start to converge (see Fig.~\ref{fig:ITP-IEV-3D}). Fig.~\ref{fig:ITP-IEV-3D} also shows that the size of the eigenvector basis strongly depends on the strength of the boson-fermion attraction. Of course, due to the spatial complexity, the ITP-IEV-3D method does not always allow to reach the required accuracy.
On the other hand, in the case of the ITP-IEV-1D method, we can use very large eigenvector bases and improve the results compared to the ITP-IEV-3D method, meaning both the ground state energy of the Bose-Fermi mixture and the bosonic and fermionic densities. This is clearly seen in Tab.~\ref{tab:energies} and Fig.~\ref{fig:dens_g1_g3_all} for $g_{BF} = -3$, where we used $2024$ basis eigenvectors. However, the ITP-IEV-1D requires a longer simulation time for obvious reasons (which is the need to construct three-dimensional basis functions from the set of one-dimensional harmonic oscillator wave functions). Note that Tab.~\ref{tab:table3} (second row) shows the wall clock times for the simulations performed within the ITP-IEV-1D method and the basis consisted of $969$ eigenvectors. For the $2024$ large eigenvector basis (used to determine the ground state energy and the bosonic and fermionic densities) the elapsed times are significantly longer.

\begin{figure}[bht]
\centering
\begin{subfigure}{.5\textwidth}
  \centering
  \includegraphics[width=.8\linewidth] {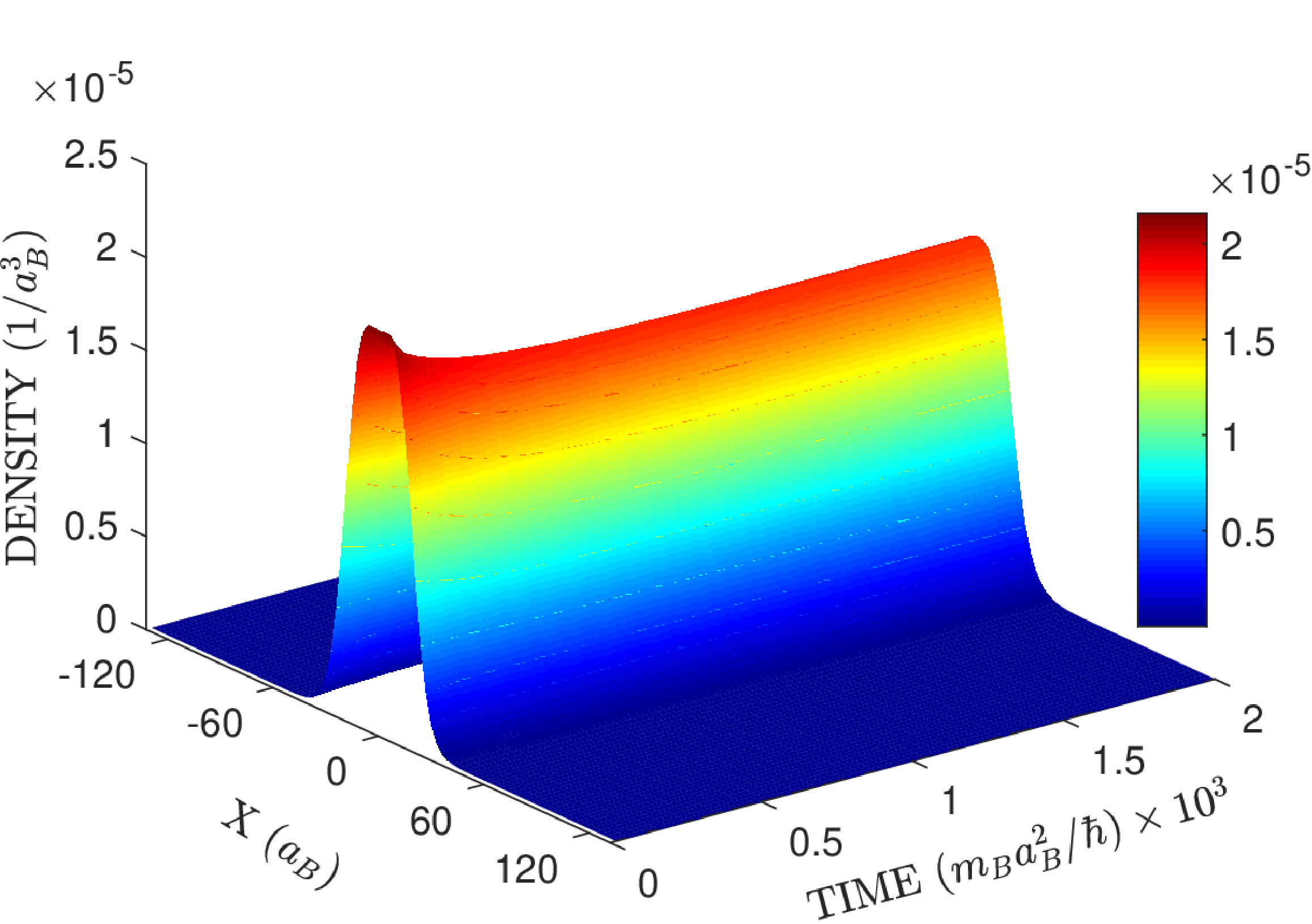}   
\vspace{1cm}
\end{subfigure}%
\begin{subfigure}{.5\textwidth}
  \centering
  \includegraphics[width=.8\linewidth] {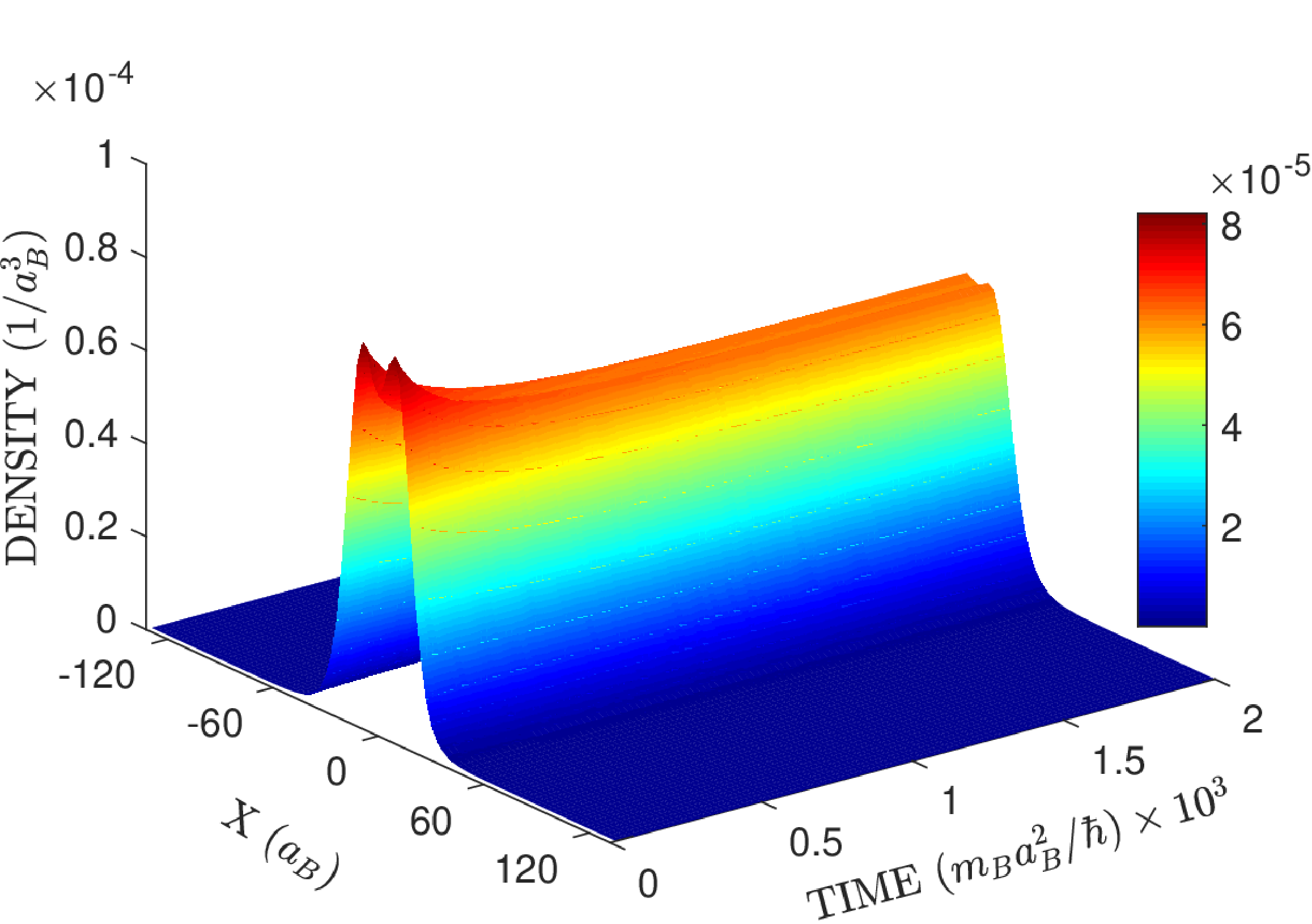}    
\vspace{1cm}
\end{subfigure}
\begin{subfigure}{.5\textwidth}
  \centering
  \includegraphics[width=.8\linewidth]{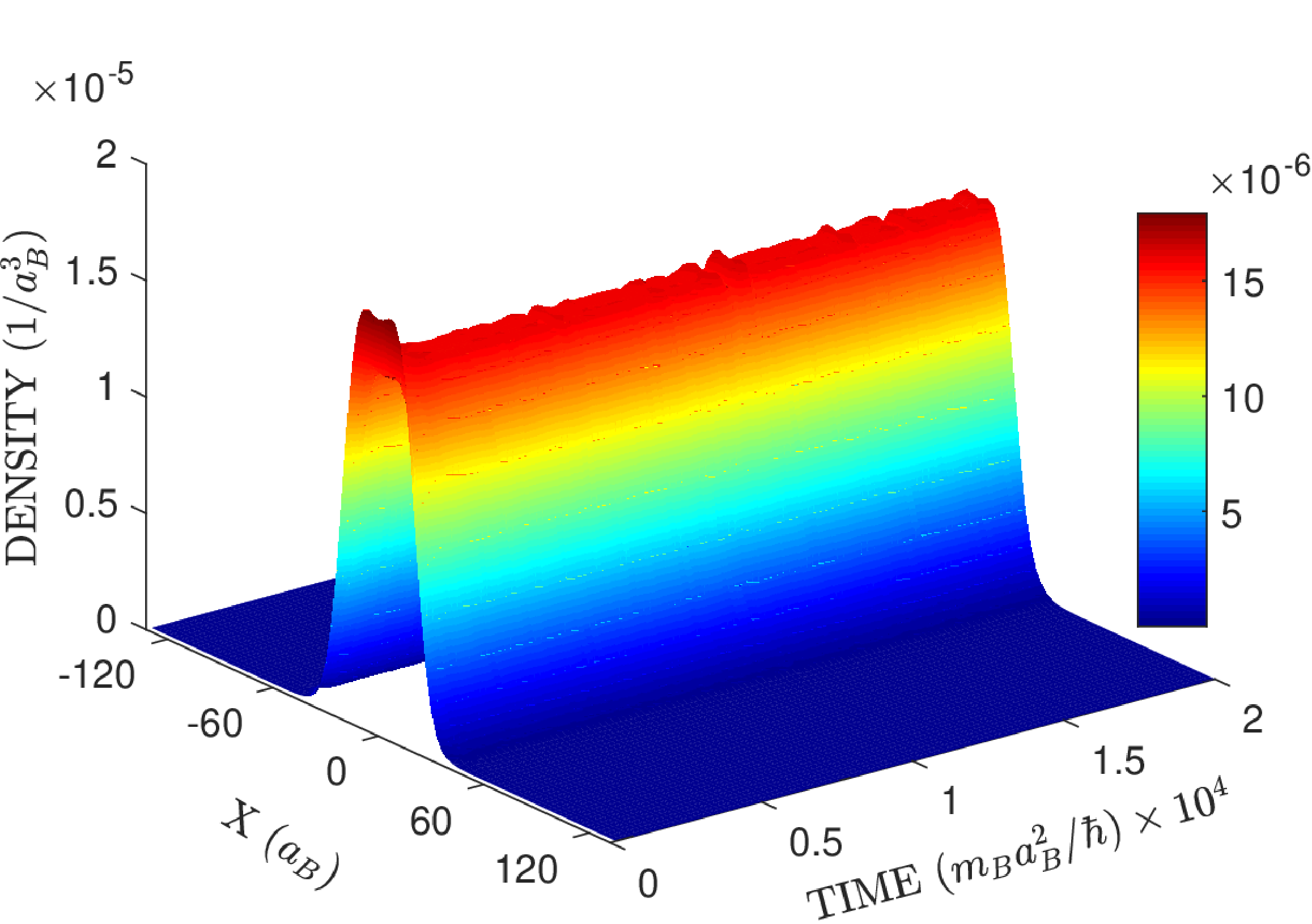}		
\end{subfigure}%
\begin{subfigure}{.5\textwidth}
  \centering
  \includegraphics[width=.8\linewidth]{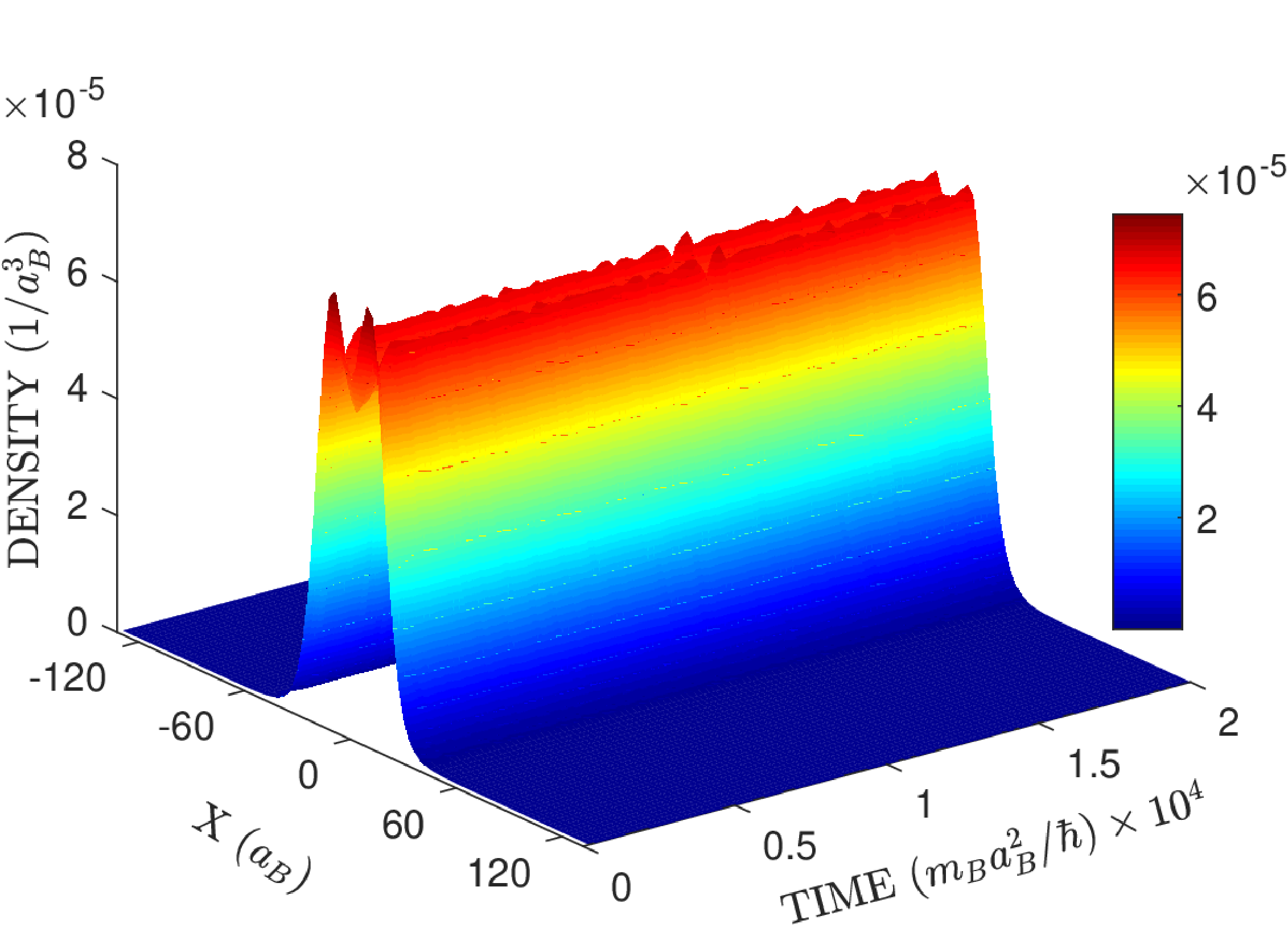}	
\end{subfigure}
\caption{Bosonic and fermionic densities after the trap is removed instantaneously. (Top) Figures represent bosonic (left) and fermionic (right) densities obtained within the ITP-ITP-GS method. The Bose-Fermi interaction parameter is equal to $g_{BF} = -4$. (Bottom) Figures represent bosonic (left) and fermionic (right) densities in the case of the A-RTP method and the interaction parameter $g_{BF} = -5$. The system consists of $N_B = 40$ bosons and $N_F = 4$ fermions. The droplet formation is clearly visible in both cases.}
\label{fig:Droplet_ITP-ITP-GS}
\end{figure}

\subsection{Self-bound Bose-Fermi droplets}

When the attraction between bosons and fermions is appropriately tuned, we can observe the formation of self-bound quantum droplets. The droplets are stabilized by the higher-order term in the Bose-Fermi coupling (see Eq.~\ref{IntEneBosFer}). They are similar in nature to experimentally observed quantum droplets in a dilute gas of dysprosium atoms, which have the largest dipolar magnetic moment of all atomic species, and in a mixture of two Bose-Einstein condensates of different species.

To observe droplet formation, we first generate the ground state of the Bose-Fermi mixture in the presence of the trap. Then we switch off the trapping potential and run the simulations for a while. If the Bose-Fermi attraction is too weak, no quantum droplets will form and the mixture will simply melt away. However, if the attraction is strong enough, the formation of a quantum droplet is clearly visible, see Fig.~\ref{fig:Droplet_ITP-ITP-GS} (top panel) for the case $g_{BF} = -4$. Similar results are obtained for the A-RTP method (Fig.~\ref{fig:Droplet_ITP-ITP-GS}, bottom panel, for $g_{BF} = -5$). Note that, in addition to the Friedel oscillations, small oscillations are present in the fermionic and bosonic densities in the case of the A-RTP method. These are the tiny oscillations in the ground state solution that are amplified by the sudden removal of the trapping potential. Although Fig.~\ref{fig:Droplet_ITP-ITP-GS} shows the bosonic and fermionic densities for a droplet consisting of a few fermions, it is worth mentioning that arbitrarily large Bose-Fermi droplets exist, as indicated  in Ref. \cite{Rakshit18}, based on general considerations and the hydrodynamic model.

Quantum self-bound Bose-Fermi droplets are a novel form of matter. They are promising for the study of a variety of physical phenomena related to polaron physics, boson-mediated pairing, and fermionic superfluidity. The stabilization mechanism involving the Fermi pressure brings some analogies to astronomical objects such as white dwarfs. Then the dynamics of Bose-Fermi droplets could simulate some astronomical processes, such as the tidal stripping of a white dwarf star orbiting a black hole \cite{Karpiuk21,Nikolajuk25}.

\section{Conclusions}\label{Conclusions}

In summary, we have presented different numerical approaches to the ground state properties (energies and densities) of Bose-Fermi mixtures and discussed their efficiency in the range of parameters where the formation of self-bound quantum droplets is expected. Both the wall clock time and the RAM memory requirements are compared. The comparison is made when two criteria are satisfied, monitoring the change in the total energy of the system and the change in the bosonic and fermionic densities over the entire numerical grid. Our numerical results clearly favor the method combining imaginary time propagation with Gram-Schmidt orthonormalization, i.e. the ITP-ITP-GS method. It provides the best estimation of the ground state energy of the Bose-Fermi mixture while ensuring the shortest wall clock time for the simulations (see Tabs.~\ref{tab:energies} and \ref{tab:table3}).

\appendix

\section{Compiler options}
\label{compiler}

All codes are written in C and compiled using the gnu gcc compiler with the following options:  \\
-std=gnu99\, -Ofast\, -march=core-avx-i\, -fopenmp\, -Wall\, -Wconversion










\end{document}